\documentclass[12pt]{elsart}
\usepackage{amssymb,latexsym,amsmath,graphicx,subfigure}
\usepackage{amscd}
\usepackage{cite}
\renewcommand{\phi}{\varphi}

\begin{document}
\begin{frontmatter}
\title{Asymptotically isochronous systems}
\thanks[CA]{Corresponding author}
\author[UR,INFN]{Francesco Calogero}\ead{francesco.calogero@uniroma1.it, francesco.calogero@roma1.infn.it}, \author[UCM]{David G\'omez-Ullate\thanksref{CA}}\ead{david.gomez-ullate@fis.ucm.es}
\address[UR]{Dipartimento di Fisica, Universit\`{a} di Roma ``La Sapienza'', 00185
Rome, Italy}
\address[INFN]{Istituto
Nazionale di Fisica Nucleare, Sezione di Roma}
\address[UCM] { Departamento de F\'isica Te\'orica II, Universidad Complutense,\\ 28040
Madrid, Spain}
\begin{abstract}
Mechanisms are elucidated underlying the existence of dynamical systems
whose \textit{generic} solutions approach \textit{asymptotically} (at large
time) \textit{isochronous} evolutions: \textit{all }their dependent
variables tend \textit{asymptotically} to functions \textit{periodic} with
the \textit{same} fixed period. We focus on two such mechanisms, emphasizing
their generality and illustrating each of them via a representative example.
The first example belongs to a recently discovered class of \textit{%
integrable} indeed \textit{solvable} many-body problems. The second example
consists of a broad class of (generally \textit{nonintegrable}) models
obtained by deforming appropriately the well-known (\textit{integrable} and
\textit{isochronous}) many-body problem with inverse-cube two-body forces
and a one-body linear ("harmonic oscillator") force.

\end{abstract}

\end{frontmatter}

\section{Introduction}

Over the last three-four decades major progress occurred in the discovery
and understanding of \textit{integrable} dynamical systems with a finite or
infinite number of degrees of freedom, and over the last decade the
possibility was noticed and exploited to identify and investigate many
\textit{isochronous} dynamical systems characterized by a time evolution
\textit{completely periodic }(i. e., periodic in \textit{all} degrees of
freedom) with the \textit{same} period. This \textit{isochronous} evolution
might prevail in the \textit{entire} (natural) phase space of the model
under consideration (one talks then of an \textit{entirely isochronous}
system), implying of course that such a model is certainly \textit{integrable%
}; or it might only prevail in an \textit{open} (hence fully dimensional)
region of its (natural) phase space, a phenomenology now known to
characterize large families of \textit{nonintegrable} dynamical systems
possibly featuring quite complicated ("chaotic") behaviors outside the
\textit{isochronous }phase space region (for a review of these developments,
see \cite{C2006,C2007} ). In the present paper we discuss another,
perhaps more interesting, phenomenology, namely dynamical systems whose
\textit{generic} solutions approach \textit{asymptotically} (at large time)
\textit{isochronous} evolutions: \textit{all }their dependent variables tend
\textit{asymptotically} to functions \textit{periodic} with the \textit{same}
fixed period. The \textit{definition} of such dynamical systems is provided
by the simultaneous validity of the two formulas
\begin{subequations}
\label{Limit}
\begin{equation}
\underset{t\rightarrow +\infty }{\lim }\left[ z_{n}\left( t\right) -\tilde{z}%
_{n}\left( t\right) \right] =0~,~\ ~n=1,...,N~,  \label{Limita}
\end{equation}%
\begin{equation}
\tilde{z}_{n}\left( t+\tilde{T}\right) =\tilde{z}_{n}\left( t\right)
~,~~~n=1,...,N~.  \label{Limitb}
\end{equation}%
\textit{Notation}: the $N$ (generally \textit{complex}; but see below)
numbers $z_{n}\left( t\right) $ denote the $N$ dependent variables of the
dynamical system under consideration; we restrict consideration to the case
when $N$ is a \textit{finite} positive integer; the \textit{real} variable $%
t $ denotes the time; the $N$ functions $\tilde{z}_{n}\left( t\right) $
characterize the asymptotic behavior of the dynamical system via (\ref%
{Limita}) and the periodicity requirement (\ref{Limitb}) they satisfy
characterizes the property of \textit{asymptotic isochronicity}. This
property is supposed to hold in an \textit{open} (hence fully dimensional)
region of the phase space of the dynamical system under consideration
(possibly coinciding with its entire natural phase space): hence the
dependent variables $z_{n}\left( t\right) $ denote here (the $N$ components
of) a \textit{generic} solution of the dynamical system evolving (at least
for sufficiently large time) within that region, while the functions $\tilde{%
z}_{n}\left( t\right) ,$ which shall generally be different for different
solutions $z_{n}\left( t\right) ,$ are required to satisfy the periodicity
property (\ref{Limitb}) with the \textit{fixed} period $\tilde{T}$ (the
\textit{same} for all the solutions in the phase space region under
consideration). Of course the formula (\ref{Limita}) does not define
uniquely -- for a given $N$-vector $\underline{z}\left( t\right) $ -- a
corresponding $N$-vector $\underline{\tilde{z}}\left( t\right) $: the
time-dependent $N$-vector $\underline{\tilde{z}}\left( t\right) $ is only
identified by (\ref{Limita}) up to arbitrary corrections whose effects
disappear in the asymptotic limit $t\rightarrow \infty $. The property of
\textit{asymptotic isochronicity} is guaranteed provided there exist just
one $N$-vector $\underline{\tilde{z}}\left( t\right) $ satisfying \textit{%
both} relations (\ref{Limit}), for every \textit{generic} solution $%
\underline{z}\left( t\right) $ in an \textit{open}, fully dimensional,
region of phase space -- namely for every solution $\underline{z}\left(
t\right) $ in that region of phase space, except possibly for some \textit{%
exceptional}, generally \textit{singular}, solutions belonging to a \textit{%
lower dimensional} sector of that phase space region.

The elementary idea underlying the identification of large classes of such
\textit{asymptotically isochronous} dynamical systems is to start from
\textit{isochronous} systems and then modify them by introducing a
deformation whose effects are significant through the time evolution yet
disappear at large time: so that the modified systems loose their \textit{%
isochronous} character (at finite times) but in some sense retain it (at
large times) as the dominant feature characterizing their \textit{asymptotic}
behavior.

There are several possible ways to implement this strategy in order to
manufacture \textit{asymptotically isochronous} systems: some are rather
trivial, some less so. This kind of judgement is of course subjective: for
instance we tend to think that an important requirement for such systems to
be deemed ``interesting'' is that they be \textit{autonomous} -- because the
interest of dynamical systems is also related to their potential usefulness
in order to model natural phenomena, which are generally described by
\textit{autonomous} evolution equations -- and moreover because the freedom
to introduce instead an \textit{explicit} time dependence in the equations
of motion of a dynamical system would provide too easy a way to influence
more or less at will the asymptotic behavior of such a system. But of course
the difference between \textit{autonomous} and \textit{nonautonomous}
systems is unessential, since any \textit{nonautonomous} system can be made
\textit{autonomous} by treating time itself as an additional dependent
variable.

In this paper we focus on two mechanisms yielding (autonomous) \textit{%
asymptotically isochronous} systems, and illustrate each of them via a
representative example. The first example (see Section 2) belongs to a
recently discovered class of \textit{integrable} indeed \textit{solvable}
many-body problems \cite{CG2007b}; in this case we eventually focus on as
simple and specific an example as possible, which is also suitable to
exhibit some numerical results -- but we trust our presentation is adequate
to illustrate the generality of the approach. In this case the periodic
behavior prevailing asymptotically corresponds to a special solution of the
dynamical system under consideration belonging to a region of phase space
with \textit{positive} codimension -- albeit \textit{not} an \textit{isolated%
} solution of this system, so not quite identifiable as a \textit{limit cycle%
}. Hence this model might be considered a representative example of a
phenomenology characterized by the presence of some kind of friction. The
second example (see Section 3) consists of a broad class of models obtained
by deforming appropriately the well-known (see for instance \cite{C2001})
\textit{integrable} and \textit{isochronous} one-dimensional many-body
problem with inverse-cube two-body forces and a one-body linear (``harmonic
oscillator'') force; the alert reader will again appreciate the generality of
the approach, even though we illustrate it by focusing on a specific model
(also restricting consideration to \textit{real} dependent variables). In
this second case the time-dependent $N$-vector to which the solutions of the
model tend \textit{asymptotically} is \textit{not} restricted to be in a
sector of phase space with \textit{positive} codimension and is generally
\textit{not} itself a solution of the \textit{asymptotically isochronous} $N$%
-body model, so this phenomenology does not correspond to what is generally
referred to as a \textit{limit cycle} behavior. In each of these two cases
we back the qualitative understanding of the origin of the relevant
phenomenology with a \textit{proof} of its actual emergence, see (\ref{Limit}%
). A section entitled ``Outlook'' in which we elaborate tersely on the
generality of this phenomenology concludes the paper.

\bigskip

\section{An asymptotically isochronous class of solvable many-body problems}

A particular mechanism to manufacture \textit{integrable}, indeed \textit{%
solvable}, dynamical systems interpretable as many-body problems inasmuch as
they are characterized by Newtonian equations of motion (``acceleration equal
force'') was introduced about three decades ago \cite{C1978} and has been
subsequently exploited to identify and investigate several such systems (for
reviews of these developments see for instance \cite{C2001,C2007}).
The idea is to exploit the \textit{nonlinear} relation among the $N$
coefficients $c_{m}\left( t\right) $ of a (for definiteness, monic)
time-dependent polynomial of degree $N$ and its $N$ zeros $z_{n}\left(
t\right) $:
\end{subequations}
\begin{equation}
\psi \left( z,t\right) =z^{N}+\sum_{m=1}^{N}c_{m}\left( t\right)
z^{N-m}=\prod\limits_{n=1}^{N}\left[ z-z_{n}\left( t\right) \right] ~.
\label{psi}
\end{equation}%
A class of such systems is characterized by the fact that the $N$
coefficients $c_{m}\left( t\right) $ evolve in time according to a system of
\textit{linear} second-order \textit{constant-coefficient} ODEs, the
solution of which is a purely algebraic task (requiring essentially the
diagonalization of an explicitly known matrix of order $N$). The
determination of the corresponding time evolution of the $N$ zeros $%
z_{n}\left( t\right) $ is therefore as well a purely algebraic task:
computing the $N$ zeros of a known polynomial. And it so happens that in
many cases \cite{C1978,C2001,C2007}  this time evolution is
indeed interpretable as that characterizing a Newtonian $N$-body problem --
hence a \textit{solvable} $N$-body problem, since its solution can be
achieved by purely algebraic means.

Indeed the solution $z_{n}\left( t\right) $ of such a model is reduced to
finding the $N$ zeros of a polynomial of degree $N$ in the (\textit{complex}%
) variable $z$, see (\ref{psi}), whose coefficients $c_{m}\left( t\right) $
generally evolve exponentially in time, typically%
\begin{equation}
c_{m}\left( t\right) =\sum_{\ell =1}^{N}\left\{ \gamma ^{\left( \ell
,+\right) }u_{m}^{\left( \ell ,+\right) }\exp \left[ \lambda ^{\left( \ell
,+\right) }t\right] +\gamma ^{\left( \ell ,-\right) }u_{m}^{\left( \ell
,-\right) }\exp \left[ \lambda ^{\left( \ell ,-\right) }t\right] \right\} ,
\label{cmt}
\end{equation}%
where the $2N$ constants $\gamma ^{\left( \ell ,\pm \right) }$ are arbitrary
(to be determined by the initial data $z_{n}\left( 0\right) ,\dot{z}%
_{n}\left( 0\right) $ in the context of the initial-value problem for the $N$%
-body system) and the $2N$ numbers $\lambda ^{\left( \ell ,\pm \right) }$
respectively the quantities $u_{m}^{\left( \ell ,\pm \right) }$ are the
eigenvalues respectively the (components of the) eigenvectors of the matrix
eigenvalue problem characterizing, as explained above, the dynamics of this
system. Note that these eigenvalues and eigenvectors are associated to the
dynamical problem under consideration: they do \textit{not} depend on the
initial data identifying a particular solution, namely they are the \textit{%
same} for all the solutions of the system.

It is now clear (and indeed well known \cite{C1978,C2001,C2007}%
) that if the $2N$ eigenvalues $\lambda ^{\left( \ell ,\pm \right) }$ are
all \textit{integer} multiples of a single \textit{imaginary} number $%
i\omega $ (with $\omega >0$), $\lambda ^{\left( \ell ,\pm \right)
}=ik^{\left( \ell ,\pm \right) }\omega $ with the $2N$ numbers $k_{\ell
}^{\left( \pm \right) }$ arbitrary \textit{integers} (positive or negative,
but not vanishing), then the polynomial $\psi \left( z,t\right) $ is clearly
\textit{periodic} with the (possibly nonprimitive) period
\begin{subequations}
\begin{equation}
T=\frac{2\pi }{\omega }~,  \label{T}
\end{equation}%
\begin{equation}
\psi \left( z,t+T\right) =\psi \left( z,t\right) ~,
\end{equation}%
hence all its zeros $z_{n}\left( t\right) $ are as well \textit{periodic}
with this same period or possibly with a (generally small \cite{GoSo}) \textit{%
integer} multiple $p$ of this period, $\tilde{T}=pT,$ due to the possibility
that they exchange their role through the time evolution. Hence the
corresponding $N$-body problem is \textit{isochronous}.

And it is as well plain that if, out of the $2N$ eigenvalues $\lambda
^{\left( \ell ,\pm \right) }$, only a (nonempty) subset have the property
indicated above while \textit{all} the others feature a \textit{negative}
real part, then the many-body problem in question is \textit{asymptotically
isochronous}. This observation is not new, see for instance Section 4.2.3 of
Ref. \cite{C2001} (entitled ``Some special cases: models with a limit cycle,
models with confined and periodic motions, Hamiltonian models,
translation-invariant models, models featuring equilibrium and spiraling
configurations, models featuring only completely periodic motions''); but, to
the best of our knowledge, this mechanism yielding \textit{asymptotically
isochronous} many-body problems was never analyzed in explicit detail
(including the display of numerical results). This is what we do in this
section, by focusing on a specific model whose \textit{integrable}, indeed
\textit{solvable}, character has been ascertained only quite recently \cite%
{CG2007b}.

\bigskip

\subsection{A specific example}

This $N$-body problem (with $N\geq 3$) is characterized by the Newtonian
equations of motion
\end{subequations}
\begin{subequations}
\label{EqMot1}
\begin{eqnarray}
\ddot{z}_{n}=-a_{1}\dot{z}_{n}+a_{2}z_{n}\frac{z_{n}^{2}-5}{z_{n}^{2}-1}%
-2a_{3}\frac{z_{n}^{2}+1}{z_{n}^{2}-1}-2a_{4}z_{n} &&  \notag \\
+2\sum_{m=1,m\neq n}^{N}\frac{\dot{z}_{n}\dot{z}_{m}+a_{2}+a_{3}z_{n}+a_{4}%
\left( z_{n}^{2}-1\right) }{z_{n}-z_{m}}~,~~~n=1,...,N~, &&  \label{EqMot1a}
\end{eqnarray}%
where the $4$ ``coupling constants'' $a_{j}$ are \textit{a priori arbitrary}
complex numbers, superimposed dots denote time-differentiations and the rest
of the notation is self-evident. The \textit{solvable} character of this $N$%
-body problem hinges \cite{CG2007b} upon the following $4$ restrictions on
its \textit{initial} data:%
\begin{equation}
\sum_{n=1}^{N}\frac{1}{z_{n}\left( 0\right) \pm 1}=0~,~~~\sum_{n=1}^{N}\frac{%
\dot{z}_{n}\left( 0\right) }{\left[ z_{n}\left( 0\right) \pm 1\right] ^{2}}%
=0~,  \label{CompCond}
\end{equation}%
which are then sufficient \cite{CG2007b} to guarantee that, throughout the
time evolution,%
\begin{equation}
\sum_{n=1}^{N}\frac{1}{z_{n}\left( t\right) \pm 1}=0~,  \label{EqMot1c}
\end{equation}%
implying that for this model it is justified to assume that only the
evolution of $N-2$ particles is determined by the Newtonian equations of
motion (\ref{EqMot1a}), while the evolution of the remaining two is
determined by these conditions, see (\ref{EqMot1c}).

Then the evolution of the $N$ ``particle coordinates'' $z_{n}\left( t\right) $
-- taking generally place in the \textit{complex }$z$-plane -- coincides
with the evolution of the $N$ zeros of a monic polynomial of degree $N$ in
the variable $z$ analogous to $\psi \left( z,t\right) ,$ see (\ref{psi}),
but more specifically reading as follows \cite{CG2007b}:
\end{subequations}
\begin{subequations}
\label{PSI}
\begin{equation}
\psi \left( z,t\right) =\pi _{N}\left( z\right) +\sum_{m=1}^{N-3}\left[
c_{m}\left( t\right) \pi _{N-m}\left( z\right) \right] +c_{N}\left( t\right)
~,
\end{equation}%
\begin{equation}
\pi _{m}\left( z\right) =z^{m}-\varepsilon _{m}\frac{m}{2}z^{2}-\varepsilon
_{m+1}mz,~~~m=0,1,...,N~,  \label{X2d}
\end{equation}%
\begin{equation}
\varepsilon _{m}=1\text{\textit{\ }if }m\text{ is \textit{even~}}%
,~~~\varepsilon _{m}=0\text{ if }m\text{ is \textit{odd}~.}  \label{epsi}
\end{equation}%
And the coefficients $c_{m}\left( t\right) $ evolve indeed according to
formulas analogous to (\ref{cmt}), but more specifically reading as follows
\cite{CG2007b}:
\end{subequations}
\begin{subequations}
\begin{eqnarray}
c_{m}\left( t\right) &=&\sum_{\ell =1,\ell \neq N-1,N-2}^{N}\left\{ \gamma
^{\left( \ell ,+\right) }u_{m}^{\left( \ell ,+\right) }\exp \left[ \lambda
^{\left( \ell ,+\right) }t\right] +\gamma ^{\left( \ell ,-\right)
}u_{m}^{\left( \ell ,-\right) }\exp \left[ \lambda ^{\left( \ell ,-\right) }t%
\right] \right\} ~,  \notag \\
m &=&1,...,N-3\text{ \thinspace \thinspace and \thinspace \thinspace }m=N~,
\label{cm}
\end{eqnarray}%
\begin{eqnarray}
\lambda ^{\left( \ell ,\pm \right) } &=&\frac{-a_{1}\pm \Delta _{\ell }}{2}%
~,~~~\Delta _{\ell }^{2}=a_{1}^{2}+4\ell \left[ a_{2}+\left( 2N-\ell
-3\right) a_{4}\right] ~,  \notag \\
\ell &=&1,...,N-3~,N~.
\end{eqnarray}%
Note that the coupling constant $a_{3}$ does not appear explicitly in these
formulas, but of course all $4$ coupling constants $a_{j}$ do play a role in
determining the quantities $u_{m}^{\left( \ell ,\pm \right) }$ appearing in
the right-hand side of (\ref{cm}).

We now restrict attention to the $N=3$ case, since this is sufficient,
indeed convenient, for exhibiting quite explicitly an \textit{asymptotically
isochronous }model. Then the only relevant coefficient (see (\ref{cm})) is
\end{subequations}
\begin{subequations}
\label{c3}
\begin{equation}
c_{3}\left( t\right) =\gamma _{+}\exp \left( \lambda _{+}t\right) +\gamma
_{-}\exp \left( \lambda _{-}t\right) ~,  \label{c3t}
\end{equation}%
\begin{equation}
\lambda _{\pm }=\frac{-a_{1}\pm \Delta }{2}~,~~~\Delta
^{2}=a_{1}^{2}+12a_{2}~,  \label{landapm}
\end{equation}%
where the somewhat simplified notation we are now using is we trust
self-explanatory (and note that in this case with $N=3$ the eigenvalues $%
\lambda _{\pm }$ only depend on the two coupling constants $a_{1}$ and $%
a_{2} $). Correspondingly, the positions of the $3$ moving particles are the
$3$ zeros $z_{n}\left( t\right) $ of the third-degree polynomial%
\begin{equation}
\psi \left( z,t\right) =\pi _{3}\left( z\right) +c_{3}\left( t\right)
=z^{3}-3z+c_{3}\left( t\right) =\prod\limits_{n=1}^{3}\left[ z-z_{n}\left(
t\right) \right] ~.  \label{psit}
\end{equation}%
Note that these $3$ zeros automatically satisfy the requirements (\ref%
{EqMot1c}), which corresponds \cite{CG2007b} to the condition that the
partial derivative of $\psi \left( z,t\right) $ with respect to $z$ vanish
at $z=\pm 1,$ $\psi _{z}\left( \pm 1,t\right) =0.$

Assume now that the two coupling constants $a_{1}$ and $a_{2}$ entail, via (%
\ref{landapm}),
\end{subequations}
\begin{subequations}
\begin{equation}
\lambda _{+}=i\omega ~,~~~\lambda _{-}=-\alpha +i\beta ~,
\end{equation}%
with $\alpha $ \textit{positive}, $\alpha >0$, $\omega $ also \textit{%
positive}, $\omega >0$ (for definiteness), and $\beta $ \textit{real} but
otherwise \textit{arbitrary}. This indeed happens provided
\end{subequations}
\begin{equation}
a_{1}=\alpha -i\left( \beta +\omega \right) ~,~~~a_{2}=\frac{\omega \left(
\beta +i\alpha \right) }{3}~.  \label{a1a2}
\end{equation}%
It is now plain that the asymptotic condition (\ref{Limita}) holds now with $%
\tilde{z}_{n}\left( t\right) $ being the three roots of the polynomial $%
z^{3}-3z+\gamma _{+}\exp \left( i\omega t\right) ,$%
\begin{equation}
z^{3}-3z+\gamma _{+}\exp \left( i\omega t\right) =\prod\limits_{n=1}^{3}
\left[ z-\tilde{z}_{n}\left( t\right) \right] ~,  \label{cubic}
\end{equation}%
which provide of course also the special solution of the model (\ref{EqMot1}%
) (with $N=3$) corresponding to initial data such that $\gamma _{-}$
vanishes (see (\ref{c3t})). And it is as well plain that the time evolution
of this polynomial is periodic with period $T$, see (\ref{T}), hence the
corresponding evolution of each of its $3$ zeros is clearly periodic with
periods $T$, $2T$ or $3T$, depending whether that zero does not ``exchange
its role'' through the motion with another zero or does so with one or with
both the other two zeros.
\begin{figure}[h]\label{fig1}
\begin{center} 
\includegraphics[width=.7\textwidth]{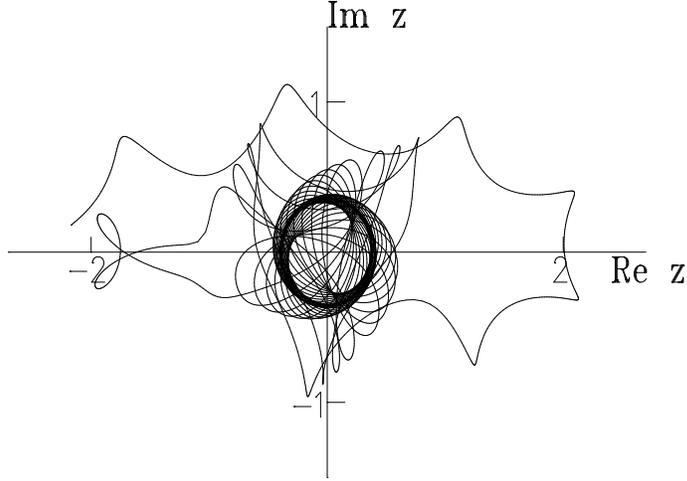}
\end{center} \caption{Trajectory of $z_1(t)$ in the complex $z$-plane from $t=0$ to $t=50$ (see text)}
\end{figure}

\begin{figure}[h]\label{fig2}
\begin{center} 
\includegraphics[width=.8\textwidth]{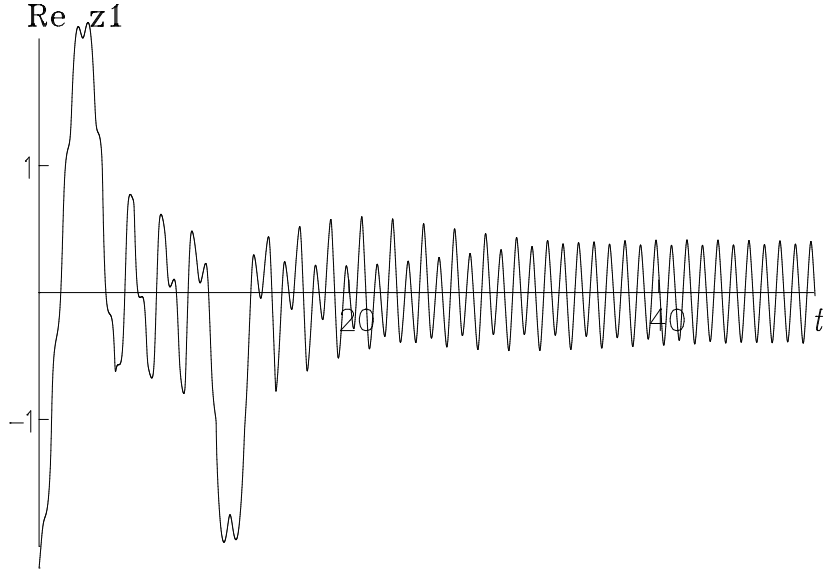}
\end{center} \caption{Plot of $\text{Re}\, z_1$ as a function of $t$ (see text)}
\end{figure}

\begin{figure}[h]\label{fig3}
\begin{center} 
\includegraphics[width=.8\textwidth]{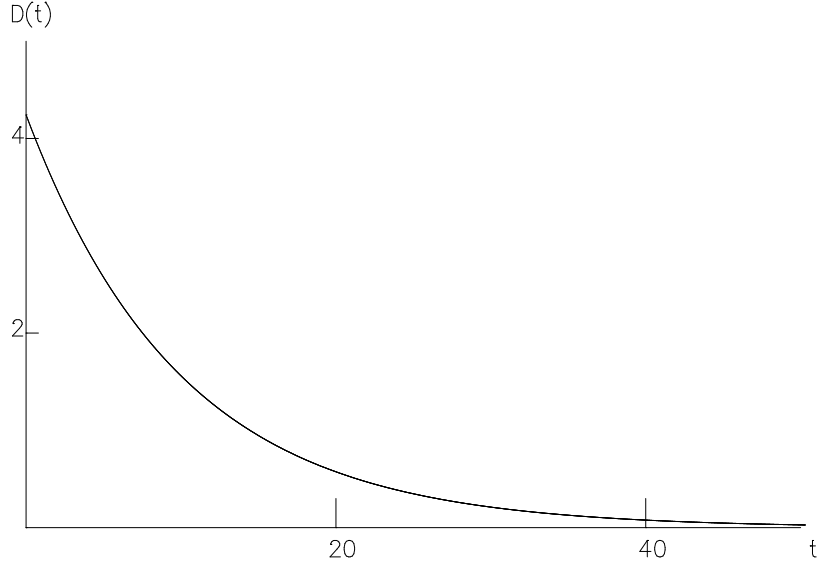}
\end{center} \caption{Plot of the distance $D(t)$ (see \eqref{Dt})}
\end{figure}
We complete this section by displaying one specific example, namely the
solution of the system of ODEs (\ref{EqMot1}) with $N=3$, $\omega =2\pi $
implying $T=1$ (see (\ref{T})), $a_{3}=a_{4}=0,$ $a_{1}$ and $a_{2}$ given
by (\ref{a1a2}) with $\alpha =0.1$ and $\beta =-3$, and with initial data

\begin{eqnarray*}
z_{1}\left( 0\right) =&-2.1702823+0.18021431 {\rm i},\quad &\dot{z}_{1}\left( 0\right) =1.2487698+0.76941297 {\rm i},\\
z_{2}\left( 0\right) =& 0.71910399-0.89149288 {\rm i}, \quad &\dot{z}_{2}\left( 0\right) =-2.7507203+1.3102500 {\rm i}, \\
z_{3}\left( 0\right) =&1.4511783+0.71127857 {\rm i}, \quad &
\dot{z}_{3}\left(0\right) =1.5019505-2.0796630 {\rm i},
\end{eqnarray*}
 satisfying the conditions (\ref{CompCond})
and entailing $\gamma _{+}=0.5+i$, $\gamma _{-}=3-3i$\thinspace (see (\ref%
{c3t})). The results displayed are, from $t=0$ to $t=50$, the trajectory of $%
z_{1}\left( t\right) $ in the complex $z$-plane (Fig. 1), the real part of $%
z_{1}\left( t\right) $ as a function of $t$ (Fig. 2) (the behavior of the
imaginary part is qualitatively analogous) and the evolution of the quantity
\begin{equation}
D\left( t\right) =\left\vert c_{3}\left( t\right) -\gamma _{+}\exp \left(
i\omega t\right) \right\vert ~~~\text{with~~~\thinspace }c_{3}\left(
t\right) =-z_{1}\left( t\right) \,z_{2}\left( t\right) \,z_{3}\left(
t\right)   \label{Dt}
\end{equation}%
(Fig. 3) that clearly provides a measure of the distance of this solution $%
\underline{z}\left( t\right) $ from its periodic limit $\underline{\tilde{z}}%
\left( t\right) $ (see (\ref{psit}) and (\ref{cubic}), as well as (\ref%
{Limit})). The numerical integration has been performed with an embedded
Runge-Kutta method of order 8(5,3) with automatic step size control, as
developed by Prince and Dormand \cite{HNW}; the integration and the
graphical output have been performed with the software \textsc{Dynamics Solver}
developed by J. Aguirregabiria.\footnote{ This software is available at
\texttt{http://tp.lc.ehu.es/jma/ds/ds.html}} The results displayed have been
obtained by integrating numerically the system of ODEs (\ref{EqMot1}),
checking throughout the integration the validity of the conditions (\ref%
{EqMot1c}) as well as the two conditions%
\begin{equation}
z_{1}\left( t\right) +z_{2}\left( t\right) +z_{3}\left( t\right)
=0~,~~~z_{1}\left( t\right) \,z_{2}\left( t\right) +z_{2}\left( t\right)
\,z_{3}\left( t\right) +z_{3}\left( t\right) \,z_{1}\left( t\right) =-3
\end{equation}%
(see (\ref{psit})). The results reported are just a representative example
of several numerical computations we did with different parameters and
initial data, computations which were found to be quite reliable and stable
unless the time evolution entailed a near collision of particles or their
passage close to the special values $z=\pm 1$ (see (\ref{EqMot1a})).

\section{A (generally nonintegrable) class of asymptotically isochronous
many-body models}

In this section we consider a class of \textit{asymptotically isochronous}
models obtained by deforming the well-known \textit{integrable} $N$-body
problem with two-body inverse cube forces and a one-body linear force, which
is of course \textit{isochronous} when no deformation is present \cite{C2001}%
. In particular we focus on the following equations of motion:
\begin{subequations}
\label{EqMot}
\begin{equation}
\ddot{x}_{n}+\frac{1}{4}\omega ^{2}x_{n}=g^{2}\sum_{m=1,m\neq n}^{N}\left(
x_{n}-x_{m}\right) ^{-3}~+F\left( w,\underline{x},\underline{\dot{x}}\right)
~,~~~n=1,...,N~,  \label{EqMota}
\end{equation}%
\begin{equation}
\dot{w}=w\left[ \alpha \log w-f\left( w,\underline{x},\underline{\dot{x}}%
\right) \right] ~,  \label{EqMotb}
\end{equation}%
with%
\begin{equation}
0<w\left( 0\right) <1~.  \label{EqMotc}
\end{equation}%
Here $N$ is an arbitrary positive integer ($N\geq 2$); the $N$ dependent
variables $x_{n}\equiv x_{n}\left( t\right) $ may be interpreted as the
coordinates of $N$ particles evolving according to the Newtonian
("acceleration equal force") equations of motion (\ref{EqMota}); these
variables $x_{n}$ are hereafter assumed to be all \textit{real} (until we
mention below to what extend the results change if the variables $x_{n}$ are
allowed to be \textit{complex}), and $\underline{x}$ denotes of course the $%
N $-vector with components $x_{n}$ (this has motivated the notational
replacement of the particle coordinates $z_{n}$ with $x_{n},$ to be kept in
mind when comparing the formulas written \thinspace in this section with
those written in the preceeding sections); likewise the auxiliary dependent
variable $w\equiv w\left( t\right) $ evolves according to the first-order
ODE (\ref{EqMotb}) with an initial condition satisfying the inequalities (%
\ref{EqMotc}) (but clearly, see below, one could replace this first-order
ODE with an \textit{appropriate} second-order "Newtonian" ODE); $t$ denotes
of course the (\textit{real}) independent variable ("time": ranging from the
\textit{initial} time $t=0$ to the \textit{asymptotic} time $t=+\infty $),
and superimposed dots denote again differentiations with respect to this
variable; $\omega ,$ $g^{2}$ and $\alpha $ are three \textit{positive} (but
otherwise \textit{arbitrary}) constants; the main restriction on the,
otherwise \textit{arbitrary}, function $F\left( w,\underline{x},\underline{v}%
\right) $ is that it vanish when $w$ vanishes,
\end{subequations}
\begin{subequations}
\begin{equation}
F\left( 0,\underline{x},\underline{v}\right) =0~,  \label{Fnzero}
\end{equation}%
and the main restrictions on the function $f\left( w,\underline{x},%
\underline{v}\right) $ is that it entail via (\ref{EqMotb}) a (very fast:
see below) asymptotic vanishing (as $t\rightarrow \infty )$ of the auxiliary
variable $w\left( t\right) $,
\end{subequations}
\begin{subequations}
\begin{equation}
\underset{t\rightarrow +\infty }{\lim }\left[ w\left( t\right) \right] =0~.
\label{wasyzero}
\end{equation}%
A condition generally sufficient (but by no means necessary) to cause this
is clearly (see (\ref{EqMotb}) with (\ref{EqMotc}) and below) the
requirement that $f\left( w,\underline{x},\underline{v}\right) $ be \textit{%
finite} and \textit{nonnegative},
\begin{equation}
0\leq f\left( w,\underline{x},\underline{v}\right) \leq a^{2}~,
\label{wfneg}
\end{equation}%
for all (\textit{real}) values of $w$, $\underline{x}$ and $\underline{v};$
it is indeed plain (for a proof, see below) that these conditions together
with (\ref{EqMotb}) entail the inequalities%
\begin{equation}
0<w\left( t\right) \leq \left[ w\left( 0\right) \right] ^{\exp \left( \alpha
t\right) }~,  \label{Inequal}
\end{equation}%
hence (see (\ref{EqMotc}) and recall that $\alpha >0$) the auxiliary
variable $w\left( t\right) $ is always positive and vanishes asymptotically
\textit{faster than exponentially},%
\begin{equation}
\underset{t\rightarrow +\infty }{\lim }\left[ w\left( t\right) \exp \left(
bt\right) \right] =0~,  \label{FastEnough}
\end{equation}%
with $b$ any arbitrary constant. Restrictions on the dependence of the
function $F\left( w,\underline{x},\underline{v}\right) $ upon the $N$%
-vectors $\underline{x}$ and $\underline{v}$ are also required: a simple
sufficient (but of course not necessary) condition, also encompassing (\ref%
{Fnzero}), is that there exist a finite (\textit{positive}) constant $C$ and
a \textit{positive} number $\beta $ such that
\end{subequations}
\begin{equation}
\left\vert F\left( w,\underline{x},\underline{v}\right) \right\vert \leq
C\left\vert w\right\vert ^{\beta }~,~~~\beta >0~,  \label{FlessC}
\end{equation}%
for all (\textit{real}) values of $w$, $\underline{x}$ and $\underline{v}$.
 Functions satisfying these conditions are for instance
\begin{eqnarray*}
F\left( w,%
\underline{x},\underline{v}\right) &=&Cw^{\beta }\left[ 1+\sum_{n=1}^{N}
\left( A_{n}^{2}\,x_{n}^{2}+B_{n}^{2}\,v_{n}^{2}\right) \right]^{-1}\\
F\left( w,\underline{x},\underline{v}\right) &=&Cw^{\beta }\exp \left[
-\sum_{n=1}^{N}\left( A_{n}^{2}\,x_{n}^{2}+B_{n}^{2}\,v_{n}^{2}\right) %
\right]
\end{eqnarray*}
 where $A_{n}$ and $B_{n}$ are arbitrary \textit{real} constants.

Our main result states that, for \textit{every} ($N$-vector) solution $%
\underline{x}\left( t\right) $ of this dynamical system, an ($N$-vector) $%
\underline{\tilde{x}}\left( t\right) $ characterizing its asymptotic
behavior (as $t\rightarrow +\infty $) via the formula (\ref{Limita}) (exists
and) has the property to be \textit{completely periodic} (i. e., \textit{%
periodic} with the \textit{same} period in \textit{all} its components), see
(\ref{Limitb}) with $\tilde{T}=T,$ see (\ref{T}). Of course this asymptotic $%
N$-vector $\underline{\tilde{x}}\left( t\right) $ will depend on the
solution $\underline{x}\left( t\right) $ under consideration -- in
particular, it will depend on the initial data, $\underline{x}\left(
0\right) $ and $\underline{\dot{x}}\left( 0\right) ,$ determining that
solution in the context of the initial-value problem for the $N$-body
problem (\ref{EqMot}): but let us re-emphasize that, for any arbitrary
choice of these data (of course, satisfying the condition $x_{n}\left(
0\right) \neq x_{m}\left( 0\right) $ for $n\neq m,$ see (\ref{EqMota})) it
shall feature the property (\ref{Limit}), namely \textit{all} solutions $%
\underline{x}\left( t\right) $ of the system (\ref{EqMot}) shall feature the
property of \textit{completely isochronous asymptotic periodicity} (\ref%
{Limit}) (with $\tilde{T}=T,$ see (\ref{T})).

This result is a natural consequence of the well-known fact (see for
instance \cite{C2001}) that \textit{all} solutions of the system of
Newtonian equations (\ref{EqMota}) \textit{without} the $F$ term in the
right-hand side are \textit{completely periodic} with period $T,$ see (\ref%
{T}), namely they \textit{all} feature themselves the property (\ref{Limitb}%
) with $\tilde{T}=T$. It stands therefore to reason that, if the function $%
F\left( w,\underline{x},\underline{v}\right) $ vanishes when $w$ vanishes,
see (\ref{Fnzero}), and if the time evolution (\ref{EqMotb}) of the
auxiliary variable $w\left( t\right) $ entails that this dependent variable
indeed vanishes asymptotically, see (\ref{wasyzero}), fast enough (see (\ref%
{FastEnough})), then \textit{asymptotically} all solutions of our model (\ref%
{EqMot}) shall behave as the solutions of the same model \textit{without}
the $F$ term, entailing the \textit{asymptotic} phenomenology (\ref{Limit})
with $\tilde{T}=T,$ see (\ref{T}).

To turn this hunch into a theorem a \textit{proof} must be provided. This we
do in the following subsection. Then in Section 4 we tersely discuss, again
in the same qualitative vein as done above, to what extent the phenomenology
described in this paper, and shown to occur in a specific, representative
model, can be expected to occur in more general contexts.

\bigskip

\subsection{A theorem and its proof}

\textit{Theorem}. The conditions (\ref{wfneg}) and (\ref{FlessC}) are
sufficient to guarantee that \textit{every} solution of the $N$-body problem
(\ref{EqMot}) with the three constants $\omega ,$ $g^{2}$ and $\alpha $ all
\textit{positive} yield the outcomes (\ref{wasyzero}) and (\ref{Limit}) with
$\tilde{T}=T,$ see (\ref{T}); in particular they guarantee that there exists,
corresponding to \textit{every} solution $\underline{x}\left( t\right) $ of
the $N$-body problem (\ref{EqMot}), an $N$-vector $\underline{\tilde{x}}%
\left( t\right) $ satisfying both formulas (\ref{Limit}) (of course, with $%
z_{n}$ replaced by $x_{n}$ and $\tilde{z}_{n}$ by $\tilde{x}_{n}$).

\textit{Proof}. First of all let us prove the inequalities (\ref{Inequal}),
obvious as they are. To this end we set
\begin{subequations}
\label{ww}
\begin{equation}
w\left( t\right) =\left[ w\left( 0\right) \right] ^{\exp \left[ \varphi
\left( t\right) \right] }~,  \label{wwt}
\end{equation}%
so that%
\begin{equation}
\varphi \left( 0\right) =0  \label{phizero}
\end{equation}%
and (from (\ref{EqMotb}))%
\begin{equation}
\dot{\varphi}\left( t\right) =\alpha +f\left[ w\left( t\right) ,\underline{x}%
\left( t\right) ,\underline{\dot{x}}\left( t\right) \right] \,\exp \left[
-\varphi \left( t\right) \right] \,\left\vert \log \left[ w\left( 0\right) %
\right] \right\vert ^{-1}~,
\end{equation}%
where we used the fact that $\log \left[ w\left( 0\right) \right]
=-\left\vert \log \left[ w\left( 0\right) \right] \right\vert ,$ see (\ref%
{EqMotc}). This ODE, together with the initial datum (\ref{phizero}) and the
inequalities (\ref{wfneg}), clearly imply that $\varphi \left( t\right) $ is
\textit{positive} and \textit{finite} for $0\leq t<\infty $, indeed validity of the
inequalities%
\begin{equation}
\alpha t<\varphi \left( t\right) <\infty ~,~~~0\leq t<\infty ~,
\end{equation}%
which, via (\ref{wwt}) and (\ref{EqMotc}), yield (\ref{Inequal}).

Next, let us introduce the counterpart of the Newtonian equations of motion (%
\ref{EqMota}), but without the $F$ term in the right-hand side:
\end{subequations}
\begin{equation}
\overset{\cdot \cdot }{\tilde{x}}_{n}+\frac{1}{4}\omega ^{2}\,\tilde{x}%
_{n}=g^{2}\sum_{m=1,m\neq n}^{N}\left( \tilde{x}_{n}-\tilde{x}_{m}\right)
^{-3}~,~~~n=1,...,N~.  \label{EqMotTilde}
\end{equation}%
Here it is justified to use the notation $\tilde{x}_{n}\equiv \tilde{x}%
_{n}\left( t\right) $ for the dependent variables, since it is well-known
\cite{C2001} that \textit{all} the solutions of this Newtonian $N$-body
problem are \textit{completely periodic} with period $T,$ see (\ref{T}),
consistently with (\ref{Limitb}) with $\tilde T =T$.

Let us now remark that, due to the strict positivity of $g^{2},$ this system
of ODEs entails that
\begin{subequations}
\label{bounds}
\begin{equation}
\left\vert \tilde{x}_{n}\left( t\right) -\tilde{x}_{m}\left( t\right)
\right\vert >\tilde{c}^{2}~,~~~~\tilde{c}^{2}>0~,~~~n\neq m~,~~~0\leq
t<\infty ~,
\end{equation}%
where $\tilde{c}^{2}$ is a time-independent constant that generally depends
on the particular solution under consideration but is certainly strictly
positive, $\tilde{c}^{2}>0$. Likewise, again due to the strict positivity of
$g^{2},$ the system of ODEs (\ref{EqMota}) with (\ref{FlessC}) and (\ref%
{Inequal}) (entailing $\left\vert F\left( w,\underline{x},\underline{v}%
\right) \right\vert \leq D,~D=C\left\vert w\left( 0\right) \right\vert
^{\beta }$) implies that%
\begin{equation}
\left\vert x_{n}\left( t\right) -x_{m}\left( t\right) \right\vert
>c^{2}~,~~~~c^{2}>0~,~~~n\neq m~,~~~0\leq t<\infty ~,
\end{equation}%
where $c^{2}$ is again a time-independent constant that generally depends on
the particular solution under consideration but is certainly strictly
positive, $c^{2}>0$. Moreover the systems of ODEs (\ref{EqMotTilde}) and (%
\ref{EqMot}) with (\ref{FlessC}) and (\ref{Inequal}) clearly imply that, for
all (finite, positive) time, the functions $\tilde{x}_{n}\left( t\right) $
and $x_{n}\left( t\right) $ are finite.

Let us now set
\end{subequations}
\begin{equation}
\xi _{n}\left( t\right) =x_{n}\left( t\right) -\tilde{x}_{n}\left( t\right)
~.  \label{zita}
\end{equation}%
These functions $\xi _{n}\left( t\right) $ satisfy -- as implied by
subtracting (\ref{EqMotTilde}) from (\ref{EqMota}) -- the system of ODEs
\begin{subequations}
\label{ODEcsi}
\begin{equation}
\ddot{\xi}_{n}+\frac{1}{4}\omega ^{2}\,\xi _{n}+g^{2}\,\sum_{m=1,m\neq n}^{N}%
\left[ \xi _{n}-\xi _{m}\right] \varphi _{nm}\left( \underline{x},\underline{%
\tilde{x}}\right) =F\left[ w,\underline{x},\underline{\dot{x}}\right] ~
\label{ODEcsia}
\end{equation}%
with
\begin{equation}
\varphi _{nm}\left( \underline{x},\underline{\tilde{x}}\right) =\frac{\left(
x_{n}-x_{m}\right) ^{2}+\left( x_{n}-x_{m}\right) \left( \tilde{x}_{n}-%
\tilde{x}_{m}\right) +\left( \tilde{x}_{n}-\tilde{x}_{m}\right) ^{2}}{\left(
x_{n}-x_{m}\right) ^{3}\left( \tilde{x}_{n}-\tilde{x}_{m}\right) ^{3}}~.
\label{ODEcsib}
\end{equation}%
Note that the above bounds, (\ref{bounds}), as well as the finiteness of $%
x_{n}$ and $\tilde{x}_{n}$ for all (positive) time, guarantee that these
functions $\varphi _{nm}\left( \underline{x},\underline{\tilde{x}}\right) $
remain \textit{finite} for all time, namely that there always exist
time-independent \textit{finite} upper and lower bounds $\varphi _{\pm }$
satisfied by them for all time,
\begin{equation}
\varphi _{-}\leq \varphi _{nm}\left( \underline{x},\underline{\tilde{x}}%
\right) \leq \varphi _{+}~.  \label{boundsphi}
\end{equation}%
These bounds depend of course on the particular solutions $\underline{x}$
and $\underline{\tilde{x}}$ under consideration, but let us re-emphasize
that, for any such solutions, they are \textit{finite}.

It is now clear that the theorem is proven if we can show that this system
of ODEs admits a solution satisfying the asymptotic condition
\end{subequations}
\begin{equation}
\underset{t\rightarrow +\infty }{\lim }\left[ \xi _{n}\left( t\right) \right]
=0~,~~~n=1,...,N  \label{asyzita}
\end{equation}%
(see (\ref{Limita}) and (\ref{zita})). As can be easily verified such a
solution of (\ref{ODEcsi}) is provided by the formula
\begin{subequations}
\begin{equation}
\xi _{n}\left( t\right) =\int_{t}^{\infty }dt^{\prime }F\left[ w\left(
t^{\prime }\right) ,\underline{x}\left( t^{\prime }\right) ,\underline{\dot{x%
}}\left( t^{\prime }\right) \right] \,G_{n}\left( t,t^{\prime }\right)
~,~~~n=1,...,N~,  \label{Solcsi}
\end{equation}%
where the functions $G_{n}\left( t,t^{\prime }\right) $ are the Green's
functions associated with the left-hand side of the system of ODEs (\ref%
{ODEcsia}), namely the solutions of the system of ODEs
\begin{eqnarray}
\frac{\partial ^{2}G_{n}\left( t,t^{\prime }\right) }{\partial t^{2}}+\frac{1%
}{4}\omega ^{2}\,G_{n}\left( t,t^{\prime }\right) &&  \notag \\
+g^{2}\,\sum_{m=1,m\neq n}^{N}\left[ G_{n}\left( t,t^{\prime }\right)
-G_{m}\left( t,t^{\prime }\right) \right] \,\varphi _{nm}\left[ \underline{x}%
\left( t\right) ,\underline{\tilde{x}}\left( t\right) \right] =0~,~~~t\leq
t^{\prime }~, &&~  \notag \\
G_{n}\left( t,t\right) =0~,~~~\left. \frac{\partial G_{n}\left( t,t^{\prime
}\right) }{\partial t}\right\vert _{t=t^{\prime }}=-1~,~~~n=1,...,N~. &&
\label{Green}
\end{eqnarray}%
Indeed, while these Green functions cannot be computed explicitly (since we
do not know the $N$-vectors $\underline{x}\left( t\right) $ and $\underline{%
\tilde{x}}\left( t\right) ,$ hence neither the functions $\varphi _{nm}\left[
\underline{x}\left( t\right) ,\underline{\tilde{x}}\left( t\right) \right] $%
), it is plain from the linear character of this system of ODEs and from the
bounds (\ref{boundsphi}) that these Green functions can grow (in modulus) at
most exponentially as $t\rightarrow \infty $ and/or $t^{\prime }\rightarrow
\infty $ ; so that the \textit{faster than exponential} vanishing of $F\left[
w\left( t^{\prime }\right) ,\underline{x}\left( t^{\prime }\right) ,%
\underline{\dot{x}}\left( t^{\prime }\right) \right] $ as $t^{\prime
}\rightarrow \infty $ (implied by (\ref{FlessC}) with (\ref{FastEnough}))
entails that the integral in the right-hand side of the solution formula (%
\ref{Solcsi}) vanishes asymptotically (as $t\rightarrow \infty $).  \qed

\textit{Remark}. It is clear how this example could have been made more
general by allowing the function $F$ appearing in the right hand side of (%
\ref{EqMota}) to depend on the index $n,$ and/or by replacing the single
auxiliary variable $w\left( t\right) $ by a $J$-vector $\underline{w}\left(
t\right) $ with $J$ an arbitrary positive integer, and so on; without
invalidating our conclusion, but complicating our proof. Let us also
re-emphasize that the hypotheses made above to prove this \textit{theorem}
are \textit{sufficient} but by no means \textit{necessary} for its validity.
More specific, and possibly considerably less stringent, conditions yielding
an analogous conclusion can and will be introduced whenever this kind of
result shall be considered in specific (possibly applicative) contexts. Our
motivation to assume here quite simple (hence overly stringent) hypotheses
is because we are just interested to show that the main idea discussed in
this paper does indeed work. \qed

\bigskip

\section{Outlook}

Clearly the kind of approaches illustrated above via the detailed treatment
of two specific examples can be applied much more widely: it will be
particularly interesting to do so in specific applicative contexts.

A natural point of departure for such applications are \textit{isochronous}
systems, namely models whose \textit{generic} solutions -- in their \textit{%
entire} natural phase space, or in \textit{open}, hence fully dimensional,
regions of it -- are \textit{completely periodic} (i. e., periodic in
\textit{all} their degrees of freedom) with the \textit{same} \textit{fixed}
period (independent of the initial data, provided they stay within the
\textit{isochronicity} region). As recently pointed out (see for instance
\cite{C2007}), quite a lot of dynamical systems can be modified so that they
become \textit{isochronous}, entailing the conclusion that \textit{%
isochronous systems are not rare}. Each of these \textit{isochronous}
systems can then be further extended -- along the lines obviously suggested
by the treatment detailed above, see in particular the specific case treated
in Section 3 -- in order to generate classes of \textit{asymptotically
isochronous} systems, namely systems featuring \textit{open}, hence fully
dimensional, regions in their natural phase space (possibly including all of
it) in which \textit{all} (or \textit{almost all}) their solutions display
asymptotically a \textit{completely periodic} behavior with the \textit{same
fixed} period, see (\ref{Limit}).\textit{\ }The technique to manufacture
such generalized systems is clearly suggested by the examples treated above:
of course these systems could be \textit{autonomous, }as the examples
treated above, or they might feature an \textit{explicit} time-dependence,
as could have been included in the system treated in Section 3 by assuming
the functions $F$ and $f$ to also feature an \textit{explicit} time
dependence (but \textit{autonomous} systems are generally more interesting
than \textit{nonautonomous} ones).

Often the natural context to investigate \textit{isochronous} systems is in
the \textit{complex} rather than the \textit{real} \cite{C2001,C2007}
-- although every system with \textit{complex} dependent variables can of
course be reformulated as a system with twice as many \textit{real}
dependent variables. Hence it may be of interest to mention how the findings
detailed in Section 3 would be affected if the dependent variables $x_{n}$
and $w$ in the model (\ref{EqMot}) were allowed to be \textit{complex} --
keeping of course \textit{real} the time $t$ and \textit{positive} the
constant $\omega $, while the constant $g^{2}$ could now also be \textit{%
complex}. It is then well known \cite{C2001, C2007}  that the \textit{%
isochronous} character of the motions still prevails for the (\textit{%
integrable} indeed \textit{solvable}) many-body problem (\ref{EqMota})
\textit{without} the $F$ term (i. e., with an identically vanishing $F;$ see
(\ref{EqMotTilde})) -- describing motions taking place in the \textit{complex%
} $z$-plane rather than on the \textit{real} line. But in the \textit{complex%
} context the \textit{isochronous} behavior is a bit different than in the
\textit{real} context: the phase space is then divided into sectors
separated by lower-dimensional manifolds characterized by solutions which
hit a \textit{singularity} at a finite time due to a particle collision; an
event forbidden in the \textit{real} case with \textit{positive} $g^{2},$
when the particles move on the \textit{real} axis and the two-body force,
\textit{singular} at zero separation, is \textit{repulsive}, see (\ref%
{EqMota}), but which can happen in the \textit{complex} case, although not
for \textit{generic} initial data. In the different sectors the motion is
still \textit{completely periodic}, but with different periods,
characterizing each sector and being (generally rather small \cite{GoSo})
\textit{integer} multiples of the basic period $T,$ see (\ref{T}).
Accordingly, the \textit{generic} solution of the (generally \textit{%
nonintegrable}) generalized model (\ref{EqMot}) will be \textit{nonsingular}
throughout its time evolution and it shall eventually settle within a
sector, approaching asymptotically one of the \textit{completely periodic}
solutions in that sector of the (\textit{integrable}) model (\ref{EqMota})
with identically vanishing $F.$

A somewhat analogous outcome obtains for the model analogous to (\ref{EqMot}%
) but with (\ref{EqMota}) replaced by
\end{subequations}
\begin{equation}
\ddot{z}_{n}+\frac{1}{4}\omega ^{2}z_{n}=\sum_{m=1,m\neq n}^{N}\left[
g_{nm}^{2}\left( z_{n}-z_{m}\right) ^{-3}\right] ~+F\left( w,\underline{z},%
\underline{\dot{z}}\right) ~,~~~n=1,...,N~,  \label{EqMotGen}
\end{equation}%
featuring $N\left( N-1\right) $ different coupling constants $g_{nm}^{2}$
acting among every particle pair. In this case the model without $F$ is
generally \textit{not integrable}, yet (if considered in the \textit{complex}%
$,$ namely without restricting the dependent variables $z_{n}$ -- nor, for
that matter, the coupling constants $g_{nm}^{2}$ -- to be \textit{real}) it
still does feature an \textit{open}, hence fully dimensional, region in its
phase space where \textit{all} solutions are \textit{completely periodic}
with the same period $T$, see (\ref{T}) \cite{C2002,C2007}; while in
other regions of its phase space it might also be \textit{periodic} but with
periods $\tilde{T}=pT$ where the numbers $p$ are \textit{integers} but might
be very large, or it might even display an \textit{aperiodic}, quite
complicated (in some sense \textit{chaotic}) behavior \cite{CS2002} (for
recent progress in the understanding of this phenomenology see \cite{CGSSa,CGSSb,FG,GS}). It then stands to reason that the
solutions of the generalized model (\ref{EqMot}) with (\ref{EqMota})
replaced by (\ref{EqMotGen}) (and of course $\underline{x}$ in (\ref{EqMotb}%
) replaced by $\underline{z}$) shall again approach asymptotically solutions
-- including, from \textit{open} regions of initial data, \textit{completely
periodic} ones -- of the model (\ref{EqMotGen}) \textit{without} $F$:
entailing a remarkable, and quite rich, phenomenology. Clearly our
motivation to mention this specific model is because of its prototypical
role: indeed, the main aspects of this phenomenology shall also characterize
the large class of \textit{isochronous} (but by no means necessarily \textit{%
integrable}) systems that can now be manufactured \cite{C2007}, once they
are extended by adding to their equations of motion other, fairly general,
terms having the property to disappear asymptotically (as $t\rightarrow
+\infty $), as a consequence of the very dynamics implied by these extended
equations of motion.

In conclusion let us re-emphasize that these results (as indeed all
mathematically correct findings) might well be deemed remarkable or trivial,
depending on the level of understanding of the reader. Once their foundation
is understood, it becomes obvious how they can be extended to many other
models -- suggesting an ample applicative potential. But these developments
exceed the scope of this paper.

\begin{ack}

One of us (FC) would like to thank Fran\c{c}ois Leyvraz for several
illuminating discussions. The research reported in this paper has profited
from visits by each of the two authors in the Department of the other
performed in the framework of the exchange program among our two
Universities. The research of DGU is supported in part by the Ram\'on y Cajal program of the
Ministerio de Ciencia y Tecnolog\'{i}a and by the DGI under grants
FIS2005-00752 and MTM2006-00478.
\end{ack}

\end{document}